\documentclass[12pt]{article}
\usepackage{rotating}
\usepackage{epsfig,amssymb,amsmath}
\usepackage{ulem}
\input{epsf}

\def\laq{\raise 0.4 ex \hbox{$<$}\kern -0.8 em\lower 0.62 ex\hbox{$\sim$}}
\def\gaq{\raise 0.4 ex \hbox{$>$}\kern -0.7 em\lower 0.62 ex\hbox{$\sim$}}

\def\beq{\begin{equation}}
\def\eeq{\end{equation}}
\def\beqa{\begin{eqnarray}}
\def\eeqa{\end{eqnarray}}

 \def\lsim{\mathrel{\rlap{\lower4pt\hbox{\hskip1pt$\sim$}}
    \raise1pt\hbox{$<$}}} \def\gsim{\mathrel{\rlap{\lower4pt\hbox{\hskip1pt$\sim$}}
    \raise1pt\hbox{$>$}}}
\def\sqr#1#2{{\vcenter{\vbox{\hrule height.#2pt
         \hbox{\vrule width.#2pt height#1pt \kern#1pt
         \vrule width.#2pt}
         \hrule height.#2pt}}}}

 \def\frac#1#2{{\textstyle{{#1}\over
{#2}}}} 
\def\lsim{\mathrel{\rlap{\lower4pt\hbox{\hskip1pt$\sim$}}
\raise1pt\hbox{$<$}}}
\def\gsim{\mathrel{\rlap{\lower4pt\hbox{\hskip1pt$\sim$}}
\raise1pt\hbox{$>$}}} \def\sqr#1#2{{\vcenter{\vbox{\hrule height.#2pt
\hbox{\vrule width.#2pt height#1pt \kern#1pt \vrule width.#2pt} \hrule
height.#2pt}}}}

\def\beq{\begin{equation}} \def\eeq{\end{equation}}
\def\beqa{\begin{eqnarray}} \def\eeqa{\end{eqnarray}}

\def\gappeq{\mathrel{\rlap {\raise.5ex\hbox{$>$}} {\lower.5ex\hbox{$\sim$}}}}

\def\lappeq{\mathrel{\rlap{\raise.5ex\hbox{$<$}}
{\lower.5ex\hbox{$\sim$}}}}

\usepackage{color}
\usepackage{cite}

\begin{document}
\pagestyle{plain}

\begin{flushright}
January 5th, 2023
\end{flushright}
\vspace{15mm}

\begin{center}

{\Large\bf Theories of gravity with nonminimal matter-curvature coupling and
the de Sitter swampland conjectures}

\vspace*{1.0cm}

Orfeu Bertolami$^{1,2}$, Cl\'audio Gomes$^{2}$ and Paulo M. S\'a$^{3,4}$ \\
\vspace*{0.5cm}
{$^{1}$ Departamento de F\'{\i}sica e Astronomia, Faculdade de Ci\^encias,
Universidade do Porto, \\
Rua do Campo Alegre s/n, 4169-007 Porto, Portugal}\\

{$^{2}$ Centro de F\'{\i}sica das Universidades do  Minho e do Porto,
Rua do Campo Alegre s/n, 4169-007 Porto, Portugal}\\

{$^{3}$ Departamento de F\'{\i}sica, Faculdade de Ci\^encias e Tecnologia,
Universidade do Algarve, \\
Campus de Gambelas, 8005-139 Faro, Portugal}\\

{$^{4}$ Instituto de Astrof\'{\i}sica e Ci\^encias do Espa\c co,
Faculdade de Ci\^encias, Universidade de Lisboa,
Campo Grande, 1749-016 Lisboa, Portugal}

\vspace*{2.0cm}
\end{center}

\begin{abstract}
\noindent
We discuss, in the context of alternative theories of gravity with
nonminimal coupling between matter and curvature, if inflationary solutions
driven by a single scalar field can be reconciled with the swampland
conjectures about the emergence of de Sitter solutions in string theory.
We find that the slow-roll conditions are incompatible with the swampland
conjectures for a fairly generic inflationary solution in such alternative
theories of gravity. 
\end{abstract}

\vfill
\noindent\underline{\hskip 140pt}\\[4pt]
\noindent
{E-mail addresses: orfeu.bertolami@fc.up.pt; claudio.gomes@fc.up.pt;
pmsa@ualg.pt}

\newpage
\section{Introduction}
\label{sec:intro}

Swampland conjectures have been proposed in order to distinguish
consistent-looking low-energy effective field theories that do not admit
a suitable ultraviolet completion in string theory --- and, therefore,
are said to be in the swampland --- from those that lie in the string
theory landscape.
This is particularly relevant as it is notoriously difficult to obtain
inflation from the fundamental fields that naturally arise in string theory. 

This difficulty is somewhat surprising as in $N=1$ supergravity 
--- which, under certain conditions, can be thought to be a low-energy limit
of string theory --- inflation can be rather easily setup (see,
for instance, Ref.~\cite{Adams}).
In fact, alternative routes to obtain inflation in string theory have been
discussed, but they tend to be more involved (see, for instance,
Ref.~\cite{KKLT}). 
It is relevant to point out that some phenomenologically viable string
models, the ones with an intermediate Grand Unified Theory energy scale,
ask for a period of inflation for its full implementation \cite{OBRoss87}.

The above-mentioned swampland conjectures are concretely a broad range of 
assumptions about the conditions required to admit local gauge symmetries and
at least one Planck mass particle so to account for the weakness of gravity.
One must also require that high-order terms in the effective action do not
admit superluminal propagation (see Ref.~\cite{Palti} for a review).
To our knowledge, there is no assumption, among this set of requirements,
concerning the Strong Equivalence Principle and 
implying that the gravity
theory is necessarily General Relativity. 

Thus, it is natural to ask if the swampland conjectures hold for
alternative theories of gravity in the context of which 
single-field inflation can take place.
This is the case of gravity theories with nonminimal coupling between
matter and curvature \cite{BBHL} where inflationary solutions can be
found \cite{GRB}.

In order to be more specific about the conditions to be met,
let us review the swampland conjectures 
relevant for our discussion.
These conjectures impose some constraints on scalar fields emerging at low
energy, generically denoted by $\phi$ \cite{OV,OOSV}, namely:
\beq
{\Delta \phi \over M_{\rm P}} < c_1,
	\label{eq:SC1}
\eeq
\beq
M_{\rm P}{|\partial_\phi V| \over V} > c_2,
	\label{eq:SC2}
\eeq
where $\Delta\phi$ is the range of variation of the field, 
$M_{\rm P}\equiv M_{\rm Pl}/\sqrt{8\pi}$ is the reduced Planck's mass,
$V(\phi)$ is the scalar field potential, 
$c_1$ and $c_2$ are constants of order one,
and we have used the notation
$\partial_\phi V \equiv \partial V/ \partial\phi$.
It has been further argued that one should consider 
the more refined condition 
\cite{OPSV,garg-krishnan,Andriot}
\beq
M_{\rm P}^2{\partial_{\phi\phi}^{2} V \over V} < -c_3, 
\label{eq:SC3}
\eeq
where $c_3$ is also a constant of order one and
$\partial_{\phi\phi}^{2} V \equiv \partial^2 V/\partial\phi^2$.

Conditions given by Eqs.~(2) and (3) can, in principle,
be compared with the onset conditions of single-field inflation which
require that the parameters for the inflaton field \cite{PDG2020}
\beq
\epsilon= {M_{\rm P}^2 \over 2} \left(
{\partial_\phi V  \over V} \right)^2
\label{eq:epsilon}
\eeq
and
\beq
\eta=M_{\rm P}^2  {\partial_{\phi\phi}^{2}
 V\over V}
\label{eq:eta}
\eeq
satisfy the slow-roll requirements $\epsilon \ll 1$ and
$|\eta| \ll 1$ at the onset of inflation, so that at the end of
inflation $\epsilon \sim |\eta| \sim1$. 

These last requirements are consistent with
constraints arising from the CMB data \cite{PDG2020}
(see Ref.~\cite{BS22} for a detailed discussion),
\beq
\epsilon <0.0044
\label{eq:ePlanck}
\eeq
and
\beq
\eta = - 0.015 \pm 0.006,
\label{eq:etaPlanck}
\eeq
whose values, clearly, do not match the requirements on $c_2$ and $c_3$. 

Actually, it can be shown that the incompatibility remains for whatever
number of scalar fields drives
inflation provided their kinetic energy terms are canonical \cite{BS22}.
However, it is possible to reconcile the swampland conjectures with
observations in the context of warm inflationary models \cite{Berera,Visinelli}
in the regime of strong dissipation for one \cite{mkr,bkr}
or more scalar fields \cite{BS22}. 

In what follows we shall consider the situation in the context of
a nonminimally coupled matter-curvature gravity theory in a single-field
inflationary setup to be specified below. 
Thus, in the next section, we shall detail the alternative gravity theory
in consideration and the associated inflationary model. We shall see that
despite the similarities between the slow-roll parameters in the
nonminimal coupled model and warm inflation, it is not possible, in the
context of the former, to satisfy the swampland conjectures. 
Finally, in section~\ref{sec:concl}, we present our conclusions.

\section{Theories of gravity with nonminimal matter-curvature coupling}

String theory itself does give origin to more complex gravitational
theories than General Relativity.
Effective models of string theory exhibit corrections to General Relativity
that include, for instance, high-order curvature terms and curvature
terms coupled with derivatives of the dilaton field
(see, for instance, Refs.~\cite{Zwiebach,BD,MT,BB1,BB2}).  

However, independently from string and quantum gravity considerations,
alternative theories of gravity are motivated as possible routes for 
addressing cosmological and astrophysical phenomena, such as the
accelerated expansion of the Universe and the flattening of the rotation
curves of galaxies, instead of resorting to dark energy and dark matter.
Well studied models include $f(R)$ gravity \cite{Capoz-2,DeFTs},
where the scalar curvature $R$ in the Einstein-Hilbert action is
replaced by a more general function, $f(R)$.
A further possibility to generalize General Relativity is to nonminimally
couple matter and curvature,  substituting the Einstein-Hilbert action 
by a more general form involving two functions of curvature
$f_1(R)$ and $f_2(R)$ \cite{BBHL}.
The function $f_1(R)$ has a role analogous to $f(R)$ gravity theory,
and the function $f_2(R)$ multiplies the matter Lagrangian density giving
rise to a nonminimal coupling between matter and geometry.
This possibility has been extensively studied in the context of dark matter
\cite{drkmattgal}, dark energy \cite{curraccel}, inflation \cite{GRB},
energy density fluctuations \cite{d}, gravitational waves \cite{e},
and the cosmic virial theorem \cite{f}.
This model has also been examined with the Newton-Schr\"odinger
approach \cite{h,i}.

Analytic extensions at $R=0$ of functions $f_1(R),f_2(R)$ were also
considered and constraints to the resulting nonminimally coupled 
gravity model have been computed through perturbations to the
perihelion precession of Mercury's orbit \cite{MPBD}.

It turns out that nonminimally coupled gravity modifies the gravitational
attraction by introducing both a fifth force of the Yukawa type and an extra
force which depends on the spatial gradient of the Ricci scalar $R$.
While the Yukawa force is typical also of $f(R)$ gravity, the existence of
the extra force is specific to nonminimally coupled gravity \cite{BBHL,BLP},
and it is an effect of the nonminimal coupling that induces a non-vanishing
covariant derivative of the energy-momentum tensor.
The arising Yukawa contribution can give origin to static solutions even
though in the absence of pressure \cite{i}.
The Yukawa contribution was also examined in the context of experiments in
deep ocean \cite{MBMGD} and through the Cassini radiometric experiment
\cite{BLM}. 

\subsection{Action, field equations and main features}

In the present work we consider theories of gravity with an action
functional of the form \cite{BBHL}
\beq\label{action-funct}
S=\int d^4x \sqrt{-g} \left[ {M_\text{P}^2\over2}
f_1(R)+ f_2(R) {\cal L}\right],
\eeq
where $f_i(R)$ (with $i=1,2$) are functions of the Ricci scalar curvature $R$,
${\cal L}$ is the Lagrangian density of matter, and $g$ is the metric
determinant.
The Einstein-Hilbert action of General Relativity is recovered by taking
$f_1(R)=R$ and $f_2(R) = 1$.

The variation of the action functional with respect to the metric
$g_{\mu\nu}$ yields the field equations
\begin{equation}
	\left( F_1+{2F_2{\cal L}\over M_\texttt{P}^2} \right) G_{\mu\nu}=
	{f_2\over M_\texttt{P}^2}T_{\mu\nu}
	+\Delta_{\mu\nu} \left( F_1+{2F_2{\cal L}\over M_\texttt{P}^2} \right)
	+ {1\over2} g_{\mu\nu} \left( f_1-F_1 R 
	-{2F_2{\cal L}\over M_\texttt{P}^2}R \right),
	\label{field-eq}
\end{equation}
where $G_{\mu\nu}$ is the Einstein tensor,
$F_i=\partial f_i/\partial R$ ($i=1,2$), and
$\Delta_{\mu\nu}\equiv
\nabla_\mu\nabla_\mu-g_{\mu\nu}\nabla_\alpha\nabla^\alpha$.

A relevant feature of nonminimally coupled gravity is that the
energy-momentum tensor of matter is not covariantly conserved.
Indeed, applying the Bianchi identities to
Eq.~(\ref{field-eq}), one obtains that
\begin{equation}
	\nabla^\mu T_{\mu\nu}={F_2\over f_2}
	\left( {\cal L} g_{\mu\nu}-T_{\mu\nu} \right)\nabla^\mu R,
	\label{covar-div-1}
\end{equation}
meaning that the nonminimal coupling cannot be ``gauged away''
by a convenient conformal transformation, being thus a distinctive feature
of the nonminimal model discussed here.

\subsection{Inflation in the nonminimally coupled theory}

As widely discussed, within General Relativity the swampland 
conjectures and the slow-roll conditions cannot be matched for single-field 
cold inflation (see, for instance, Refs.~\cite{Kinney,Riotto}).

Given that the incompatibility of the swampland conjectures with the
observations has been an object of critique from the authors of
Ref.~\cite{Linde2019} and that multi-field inflationary models show no
contradiction with the CMB features \cite{Wands}, it was logical to ask
whether the swampland conjectures would hold for many fields.
In fact, multi-field cosmological models open interesting
perspectives, for instance, for unification of 
dark matter and dark energy \cite{Sa2020a,Sa2020b,Sa2021,Sa2022}. 
Two-field inflationary models were first considered in the context of $N=1$
supergravity \cite{Ovrut} and their dynamics was scrutinized in
Refs.~\cite{OBRoss86,OB88} for a broader class of models.
In a broad context and in string theory, two-field inflationary with
different mass scales and an interaction term were considered in
Refs.~\cite{Linde94,Bento91,Bento92}.
In the context of the swampland conjectures, two-field inflationary models 
were examined in Refs.~\cite{Achucarro,Gashti}, where in Ref.~\cite{Achucarro}
non-canonical kinetic energy terms have been considered.
More recently, it has been shown that multi-field inflation cannot be made
compatible with the swampland conjectures without a significant amount of
dissipation \cite{BS22}.

In this paper, we will consider single-field cold inflation within
the nonminimally coupled theory of gravity defined by the action
functional~(\ref{action-funct}).

Assuming the Friedmann-Lema\^{\i}tre-Robertson-Walker metric 
\begin{equation}
	ds^2=-dt^2+a^2(t)dx^2,
\end{equation}
from Eqs.~(\ref{field-eq}) and (\ref{covar-div-1}) we obtain
\begin{equation}
	H^2 = {1\over 6F}
	\biggr[ {2f_2\rho\over M_\text{P}^2}
	-6H\dot{F}
	-f_1+ F R\biggr],
	\label{H-general}
\end{equation}
\begin{equation}
	-2\left( 2\dot{H}+3H^2 \right) F
	= {2f_2p\over M_\text{P}^2}
	+2\ddot{F}
	+6H\dot{F}
	+f_1 -F R,
	\label{Hdot-general}
\end{equation}
\begin{equation}
	\dot{\rho}+3H(\rho+p)= 
	-{F_2 \over f_2}(\rho+p)\dot{R},
	\label{3}
\end{equation}
where $a(t)$ is the scalar factor,
$H=\dot{a}/a$ is the Hubble parameter,
an overdot denote a derivative with respect to time $t$,
and we have introduced the notation
$F\equiv F_1+2F_2 p/M_\text{P}^2$.
In the above equations, we have also assumed that matter is represented
by an homogeneous scalar field $\phi$ with a Lagrangian density ${\cal L}=p$,
for which pressure and energy density are given by $p=\dot{\phi}^2/2-V(\phi)$
and $\rho=\dot{\phi}^2/2+V(\phi)$, respectively.

In what follows, we consider theories for which the pure gravitational
sector of the action has the Einstein-Hilbert form; more specifically,
we choose $f_1(R)=R$, implying $F_1=1$.

With this assumption, Eqs.~(\ref{H-general}) and (\ref{Hdot-general})
become
\begin{equation}
	H^2={1\over 3(M_\text{P}^4-4G^2)}
	\left[ \rho f_2(M_\text{P}^2-G)-3pf_2G
	-6H(M_\text{P}^2+2G)\dot{G}-6G\ddot{G} \right],
	\label{1}
\end{equation}
and
\begin{equation}
	\dot{H}=-{f_2(\rho+p)+2\ddot{G}\over 2(M_\text{P}^2+2G)},
	\label{2}
\end{equation}
where the notation $G\equiv p F_2$ was introduced.

Let us now choose
\begin{equation}
	f_2(R)=1+ \alpha\left( {R\over 6M_\text{P}^2} \right)^3,
\label{3}
\end{equation}
where $\alpha$ is a positive dimensionless parameter that sets the scale of the 
nonminimal coupling, which is not necessarily the Planck scale.
The cubic choice is the simplest power-law type function
which renders non-trivial solutions for the Friedmann equation.
In fact, for a linear monomial no real solutions for that equation are found,
and for the quadratic scenario the standard solution in General Relativity
is surprisingly retrieved; as for cubic and higher monomials the behaviour
is similar among each choice, up to small numerical factors \cite{GRB}.
Furthermore, we assume that inflation is quasi-exponential,
i.e., $V\gg \dot{\phi}^2$, implying $\rho\simeq -p \simeq V$.

Under these assumptions, and taking into account that
\beq
	R=6(\dot{H}+2H^2), 
\eeq
we obtain
\beq
	f_2 = 1+ {\alpha \over M_\text{P}^{6}} \left(
	8H^6+12H^4\dot{H}+6H^2\dot{H}^2+\dot{H}^3 \right),
	\label{approx-f2}
\eeq
\beq
	G = {\alpha \over M_\text{P}^{6}}\left(
	2H^4p+2H^2\dot{H}p+{\dot{H}^2\over 2}p
	\right),
	\label{approx-G}
\eeq
\beq
	\dot{G} = 
	{\alpha \over M_\text{P}^{6}}
	\bigg(
	8H^3\dot{H}p+4H\dot{H}^2p+2H^4\dot{p}
	+2H^2\dot{H}\dot{p}+{\dot{H}^2\over 2}\dot{p}
	+2H^2\ddot{H}p+\dot{H}\ddot{H}p
	\bigg),
	\label{approx-Gdot}
\eeq
\begin{align}
	\ddot{G} &=
	{\alpha \over M_\text{P}^{6}}
	\bigg(
	24H^2\dot{H}^2p
	+4\dot{H}^3p
	+16H^3\dot{H}\dot{p}
	+8H\dot{H}^2\dot{p}
	+8H^3\ddot{H}p
	+12H\dot{H}\ddot{H}p
	+4H^2\ddot{H}\dot{p}
	\nonumber
\\	&+2\dot{H}\ddot{H}\dot{p}
	+\ddot{H}^2p
	+2H^4\ddot{p}
	+2H^2\dot{H}\ddot{p}
	+{\dot{H}^2\over 2}\ddot{p}
	+2H^2H^{(3)}p
	+\dot{H}H^{(3)}p
	\bigg),
	\label{approx-Gddot}
\end{align}
where $H^{(3)}$ denotes the third derivative of $H$ with respect to time $t$.

Equation~(\ref{1}) can now be written as
\begin{align}
	3\left( M_\text{P}^2-{4\alpha H^4\over M_\text{P}^6}p \right)
	\left( M_\text{P}^2+{4\alpha H^4\over M_\text{P}^6}p \right) &\simeq
	\rho \left( 1+{8\alpha H^6\over M_\text{P}^6} \right)
	\left( M_\text{P}^2-{2\alpha H^4\over M_\text{P}^6}p \right)
	\nonumber
\\	&-3p^2\left( 1+{8\alpha H^6\over M_\text{P}^6} \right)
	{2\alpha H^4\over M_\text{P}^6}
\end{align}
or, for $M_\text{P}^8-4\alpha H^4V\neq0$,
\begin{equation}
	4\alpha V H^6+3M_\text{P}^8H^2-M_\text{P}^6V\simeq0,
	\label{eq1a}
\end{equation}
where we have taken only the first two terms of $f_2$ and the first term
of $G$ in Eqs.~(\ref{approx-f2}) and (\ref{approx-G}), respectively;
all the other terms in these equations, as well as the terms of $\dot{G}$
and $\ddot{G}$ in Eqs.~(\ref{approx-Gdot}) and (\ref{approx-Gddot}),
were neglected since they contain time derivatives of $H$ and $p$.

For the sake of simplicity, let us now introduce the dimensionless variable
\begin{equation}
	\overline{V}={V\over M_\text{P}^4}.
\end{equation}
Then, equation (\ref{eq1a}) becomes
\begin{equation}
	4\alpha \overline{V}H^6+3M_\text{P}^4H^2
	-M_\text{P}^6\overline{V}\simeq0,
\end{equation}
yielding the solution
\begin{equation}
	H^2\simeq {-1+\Big(\sqrt{\alpha \overline{V}^3}
	+\sqrt{1+\alpha\overline{V}^3}\Big)^{2/3}
	\over 
	2\sqrt{\alpha\overline{V}}\Big(\sqrt{\alpha \overline{V}^3}
	+\sqrt{1+\alpha\overline{V}^3} \Big)^{1/3}}M_\text{P}^2.
	\label{eq1b}
\end{equation}

Taking into account that the energy scale of inflation,
defined as $E_{\texttt{inf}}=V^{1/4}$, is much smaller than the
reduced Planck mass, the right-hand side of Eq.~(\ref{eq1b})
can be expanded in power series of $\overline{V}\ll 1$,
yielding
\begin{equation}
	H^2 \simeq {M_\text{P}^2\over 3}\overline{V}
	\left(1	- {4\over 27}\alpha\overline{V}^3 \right),
	\label{A}
\end{equation}
where the first term on the right-hand side is the General Relativity term
and the second one is a correction due the presence of a nonminimal coupling
between matter and curvature.

Let us now turn to Eq.~(\ref{2}).
It can be written as
\begin{equation}
	2\left( M_\text{P}^2+{4\alpha H^4\over M_\text{P}^6}p \right)\dot{H}
	\simeq -\left( 1+{8\alpha H^6\over M_\text{P}^6} \right)(\rho+p)
	\label{eq2a-aux}
\end{equation}
or, equivalently,
\begin{equation}
	\dot{H}\simeq -{(M_\text{P}^6+8\alpha H^6)\dot{\phi}^2
	\over 
	2(M_\text{P}^8-4\alpha M_\text{P}^4 H^4 \overline{V})},
	\label{eq2a}
\end{equation}
where we have used $\rho+p=\dot{\phi}^2$
and taken only the first two terms of $f_2$ and the first term
of $G$ in Eqs.~(\ref{approx-f2}) and (\ref{approx-G}), respectively;
all the other terms in these equations, as well as the terms of
$\ddot{G}$ in Eq.~(\ref{approx-Gddot}),
were neglected since they contain time derivatives of $H$ and $p$.

Using $H^2$ given by Eq.~(\ref{A})
and expanding in power series of $\overline{V}$, Eq.~(\ref{eq2a}) yields
\begin{equation}
	\dot{H} \simeq -{\dot{\phi}^2\over 2M_\text{P}^2}
	\left( 1+{20\over 27} \alpha\overline{V}^3\right),
	\label{B}
\end{equation}
where, again, the first term on the right-hand side is the General
Relativity term and the second is a correction due to the direct
matter-curvature coupling.

Note that, if in Eq.~(\ref{eq2a-aux}) we had also taken for $f_2$ the term
proportional to $\dot{H}$ and for $\ddot{G}$ the term proportional to 
$\dot{H}\dot{p}$, we would have obtained in the right-hand side of
Eq.~(\ref{B}) an extra term proportional to $\dot{\phi}^4\overline{V}^2$; 
however, since $\dot{\phi}^2/M_\text{P}^4\ll \overline{V}$,
this extra term can be neglected in comparison with the term proportional to 
$\dot{\phi}^2\overline{V}^3$.

Now, taking the time derivative of Eq.~(\ref{A})
and using Eq.~(\ref{B}) to eliminate $\dot{H}$, we obtain
\begin{equation}
	\partial_\phi V \simeq -3H\dot{\phi}
	\left( 1+{20\over 27}\alpha\overline{V}^3 \right)
	\left( 1-{16\over 27}\alpha\overline{V}^3 \right)^{-1}
\end{equation}
or, expanding in power series of $\overline{V}$,
\begin{equation}
	\partial_\phi V \simeq -3H\dot{\phi}
	\left( 1+{4\over 3}\alpha\overline{V}^3 \right).
	\label{C}
\end{equation}

Taking a second time derivative, Eq.~(\ref{C}) becomes:
\begin{equation}
	\partial_{\phi\phi}^2 V \simeq 3H^2
	\left( 1+{4\over 3} \alpha \overline{V}^3 \right)
	\left(-{\dot{H}\over H^2}-{{\ddot{\phi} \over {\dot{\phi}H}}}
         \right),
	 \label{D}
\end{equation}
where we have neglected terms proportional to
$\dot{\phi}^2 \overline{V}^2$.

Using Eqs.~(\ref{A}), (\ref{B}), (\ref{C}), and (\ref{D}),
the quantities $\dot{H}/H^2$ and $\ddot{\phi}/(\dot{\phi}H)$
can be expressed as
\begin{equation}
	{\dot{H}\over H^2}\simeq \epsilon 
	\left( 1-{44\over 27} \alpha \overline{V}^3 \right)
\end{equation}
and
\begin{equation}
	{\ddot{\phi}\over \dot{\phi}H} \simeq
	\epsilon \left( 1-{44\over 27} \alpha \overline{V}^3 \right)
	-\eta \left( 1-{32\over 27} \alpha \overline{V}^3 \right), 
\end{equation}
where the slow-roll parameters $\epsilon$ and $\eta$ are
given by Eqs.~(\ref{eq:epsilon}) and (\ref{eq:eta}).

Now, taking into account that in the slow-roll inflationary regime
$|\dot{H}|/H^2\ll1$ and $|\ddot{\phi}/(\dot{\phi}H)| \\ \ll 1$
we conclude that
\begin{equation}
	\epsilon\ll  1+{44\over 27} \alpha \overline{V}^3
	\quad \mbox{and} \quad |\eta|\ll 
	 1+{32\over 27} \alpha \overline{V}^3.
\end{equation}

Since $\overline{V}\equiv V/ M_\text{P}^4\ll 1$ and assuming for
naturalness that $\alpha={\cal O}(1)$, we conclude that in the nonminimally
coupled theory of gravity under consideration, the slow-roll parameters
satisfy the conditions $\epsilon\ll1$ and $|\eta|\ll1$.
Because these parameters are related to the constants $c_2$ and $c_3$
arising within the de Sitter swampland conjectures
[see Eqs.~(\ref{eq:SC1}) and (\ref{eq:SC2})]
through the relations 
\begin{equation}
	c_2^2<2\epsilon \quad \mbox{and} \quad c_3<|\eta|,
\end{equation}
we arrive at the conclusion that $c_2\ll 1$ and $c_3\ll 1$
during a quasi-exponential inflationary period.

Thus, we clearly see that the swampland conjectures cannot be met for
inflation in the context of theories of gravity with nonminimally coupled
matter and curvature. 

\section{Discussion and Conclusions\label{sec:concl}}

In this work, in the context of the nonminimally coupled matter-curvature
theory of gravity, we have considered the compatibility of the slow-roll
conditions of inflation and the de Sitter swampland conjectures. 

Despite the specificities of the nonminimally coupled theory and the fact
that it can lead to an inflationary regime which differs from the
one in General Relativity for the choice of the $f_2(R)$-function such as
in Eq.~(\ref{3}), we find that under quite general conditions the  requirements
for the inflaton potential are still very much controlled by the slow-roll
conditions.
Even though it is conceivable that the free parameter $\alpha$ introduced in
Eq.~(\ref{3}), which sets the impact of the nonminimal coupling,
could be greater than one, it cannot overcome the typical scale of the
inflaton potential and its smallness in comparison with the Planck scale.
Of course, for naturalness reasons, we assume that $\alpha={\cal O}(1)$.
Thus, we conclude that the de Sitter swampland conditions cannot be met in
the context of gravity theories with nonminimal coupling between
matter and curvature.  
We expect these conclusions to hold for any number of inflation fields.  

\vspace{0.5cm}

{\bf Acknowledgments}

\noindent
PMS acknowledges support from Funda\c{c}\~ao para a Ci\^encia e a
Tecnologia (Portugal) through the research grants UIDB/04434/2020
and UIDP/04434/2020.
OB and CG acknowledge support from Funda\c{c}\~ao para 
a Ci\^encia e a Tecnologia (Portugal) through the research project
CERN/FIS-PAR/0027/2021.

\end{document}